\documentclass[
superscriptaddress,
nofootinbib,
 amsmath,amssymb,
 aps,
 prd,
floatfix,
twocolumn
]{revtex4-2}

\usepackage{diagbox}
\usepackage{graphicx}
\graphicspath{{./pic/pp/}}
\DeclareGraphicsExtensions{.pdf,.png}
\usepackage{bm}  
\usepackage[utf8]{inputenc}

\pdfoutput=1

\usepackage[autostyle, english = british]{csquotes}
\MakeOuterQuote{"}

\usepackage[english]{babel}

\usepackage[format=plain,
            textfont=it]{caption}

\usepackage{graphicx}
\usepackage{adjustbox}
\usepackage{subcaption} 
\usepackage{capt-of}
\usepackage{amsmath,amssymb,amsfonts,tensor}
\usepackage{xcolor}
\usepackage{graphicx}
\usepackage{subcaption}
\usepackage{float}
\usepackage{dsfont}
\usepackage{enumitem}
\usepackage[section]{placeins}
\usepackage{empheq}
\usepackage{tikz-cd}

\usepackage{algorithm}
\usepackage[noend]{algpseudocode}

\usepackage{physics} 
\usepackage[percent]{overpic} 

\usepackage[colorlinks,citecolor=blue,linkcolor=blue,urlcolor=blue,pdfencoding=auto, psdextra]{hyperref}

\newcommand{\be}{\begin{equation}}
\newcommand{\ee}{\end{equation}}

\usepackage{algorithm}
\usepackage[noend]{algpseudocode}

\newcommand{\nn}{\nonumber}

\newcommand{\Wthree}[6]{\left(\begin{array}{ccc} #1 & #2 & #3 \\ #4 & #5 & #6 \end{array}\right)}
\newcommand{\Wfour}[9]{\left(\begin{array}{cccc} #1 & #2 & #3 & #4 \\ #5 & #6 & #7 & #8 \end{array}\right)^{(#9)}}
\newcommand{\Wsix}[6]{\left \{ \begin{array}{ccc} #1 & #2 & #3 \\ #4 & #5 & #6 \end{array}\right \} }

\usepackage[colorinlistoftodos,prependcaption,textsize=footnotesize]{todonotes}

\begin{document}

\title{Numerical approach to the Black-to-White hole transition}
\author{Pietropaolo Frisoni}
\email{pfrisoni@uwo.ca}
\affiliation{Dept.\,of Physics \& Astronomy, Western University, London, ON N6A\,3K7, Canada} 

\begin{abstract}
We outline an algorithm to numerically compute the black-to-white hole transition amplitude using the loop quantum gravity covariant formulation and the Lorentzian Engle-Pereira-Rovelli-Livine model. We apply the algorithm to calculate the crossing time of the transition in the deep quantum regime, comparing our result with previous analytical estimates of the same physical observable in the semiclassical limit. Furthermore, we show how to evaluate the crossing time analytically using an alternative approach to the one currently in the literature. This method requires much easier calculations and emphasizes that the crossing time does not depend on the extrinsic geometry of the transition.
\end{abstract}

\maketitle

{
  \hypersetup{linkcolor=black}
}


\section{Introduction}
At present, black holes seem to be perfectly described by classical general relativity, including their behavior in the strong field regime \cite{Abbott2016}. We have no reason to suspect that, beyond the black hole horizon, general relativity does not continue to provide a reasonable physical description inside it. The region where our knowledge falters is the center: we have no idea what happens to an object after it reaches the singularity. Furthermore, the distant future of a black hole is still quite a mystery. The calculation originally made by Hawking \cite{Hawking:1975vcx} shows that the black hole shrinks due to the back reaction of the Hawking radiation. The black hole should become smaller and smaller, but after this phase, nothing is known. The perturbative formulation of quantum gravity disregards non-perturbative quantum-gravitational phenomena. This is the reason why the full theory of quantum gravity is required. The possibility of black hole decay via gravitational quantum tunneling is currently one of the most intriguing hypotheses on the future of these objects \cite{Narlikar1974,Frolov:1979tu2, Frolov:1981mz2,Giddings1992b,Stephens1994,Mazur:2004,modesto2004disappearance,Ashtekar:2005cj,Mathur2005,Hayward2006,Balasubramanian:2006,Modesto2008a,Hossenfelder:2010a,frolov:BHclosed,Bambi2013,Gambini:2013qf,Bardeen2014,Mathur2015,Saueressig2015,Barcelo:2015uff,Barcelo2014b}. 

\medskip

In the last few years, considerable effort has been devoted to investigating the phenomenon using the covariant "spinfoam" formulation of loop quantum gravity \cite{article:Christodoulou_Rovelli_Speziale_Vilensky_2016_planck_star_tunneling_time, Characteristics_time_scales, Black_Hole_part_2, christodoulou2023geometry, Soltani_2023, RovelliLectures}. At the same time, there have been remarkable advances in the development of computational methods in the field. A few examples are the application of MCMC methods to investigate the semiclassical limit \cite{Spinfoam_Lefschetz_thimble} as well as the deep quantum regime \cite{Markov_chain_paper, VertexRenormalizationPaper}, the study of cuboid renormalization \cite{Steinh_cuboidal_renorm}, the introduction of effective spin foams \cite{Effective_spinfoam_1, Effective_spinfoam_2}, and the study of the EPRL amplitudes using high-performance computing \cite{Dona2018, Review_numerical_LQG, Francesco_draft_new_code, dona2022spinfoams}. One of the main reasons for developing techniques to compute EPRL spinfoam amplitudes was to investigate the black-to-white transition using computational methods \cite{Fabio_PhD_thesis, Frisoni:2023vnn}. In this sense, this paper aims to be the first direct link between these two research directions. 

\medskip

We outline an algorithm to compute the amplitude and apply it to calculate the crossing time of the transition. We estimate it numerically and analytically, modifying the boundary state with respect to the calculation of the same observable currently present in the spinfoam literature, corresponding to a different physical regime \cite{Fabio_PhD_thesis, christodoulou2023geometry, Characteristics_time_scales}. We find the same result: the crossing time scales linearly with the mass. The calculation described here is remarkably simpler and shows that the crossing time does not depend on the extrinsic curvature of the boundary geometry. Therefore, our result is in excellent agreement with the previous estimates, supplementing them with new physical information. The paper is organized as follows. Section \ref{sec:quantum_tunneling} briefly reviews the quantum tunneling hypothesis and the necessity of a full quantum gravity theory to describe it. In Section \ref{sec:Geometry}, we describe the external geometry of the process. In Section \ref{sec:Transition_amplitude}, we write the four-dimensional spinfoam amplitude of the black-to-white hole transition, and in Section \ref{sec:algorithm}, we outline the algorithm to compute it. Finally, in Section \ref{sec:crossing_time}, we evaluate the crossing time of the transition. Unless explicitly indicated otherwise, we use the Planck unit system ($c= \hbar = G = 1$) in the following.
\section{The quantum tunneling}
\label{sec:quantum_tunneling}
Regardless of what happens in the future, after the full evaporation of the black hole has occurred, it is reasonable to expect that in a distant forward time, all that remains is regular spacetime. That is, we expect that there is a spacetime with a causal structure after the end of the black hole evaporation. We know from classical GR that a collapsed star creates a horizon. The "cosmological censor" conjecture \cite{Penrose:1969pc} states that every singularity in classical GR is always hidden inside a horizon. Assuming that this conjecture is true since GR is invariant under time reversal, the opposite also turns out to be true. Therefore, even in the future, the "putative singularity" should be closed inside a horizon. To describe the process, we need a description of the external classical geometry and a quantum one, which provides information on the tunneling transition inside the black hole. 

\medskip

\begin{figure}[h]                      
\centering
\includegraphics[width=4cm]{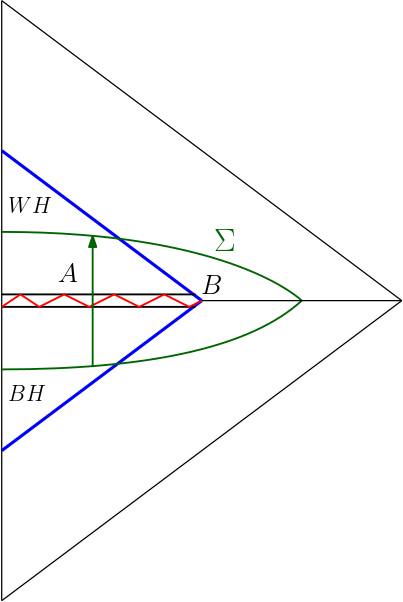}
\caption{\textit{Extended Penrose diagram describing the tunneling process. The B region represents the future of the black hole after the evaporation, whereas A is the region around the singularity center.}}
\label{fig:HR_spacetime}
\end{figure}

These ingredients are provided precisely by loop quantum gravity, which describes the transition amplitude between classical geometries. In particular, the closed surface surrounding the classical region can be arbitrarily chosen. 
Therefore, the central singularity is enclosed inside a boundary, namely a space-like surface $\Sigma$ with a classical geometry defined on it. A tunneling effect exists between the two space-like regions on "opposite sides" of the singularity. This is a purely quantum effect as these regions have no classical transition. Thus, there is no possible classical evolution. A possibility considered in the literature \cite{Rovelli2014} is that there is a white hole after the evaporation of the (small) black hole, which remains as a remnant, which should radiate in the low-frequency spectrum \cite{Kazemian_2023}. In Figure \ref{fig:HR_spacetime}, we sketch the tunneling process using an (extended) Penrose diagram. The blue lines denote the gravitational horizon, and the green ones indicate the space-like boundary surface. The B region represents the future of the black hole after the evaporation, whereas A is the region around the singularity center. The quantum theory describes the tunneling process. It is reasonable to expect that the degrees of freedom of the latter are not arbitrarily small but comparable to the black hole size. 

\medskip

First, we must truncate the theory to compute transition amplitude using covariant loop quantum gravity. We do so by discretizing $\Sigma$ and the four-dimensional interior region. If we know the (intrinsic and extrinsic) geometry of $\Sigma$, we can write an extrinsic coherent state in the Hilbert space of loop quantum gravity to the truncation.
\section{Geometry} 
\label{sec:Geometry}
The external final metric depends only on two parameters: the mass of the black hole $m$ and the "time" $T$ between the lower and upper regions \cite{Haggard_2015}. Therefore, these two parameters completely describe the external geometry. There are several characteristics and time scales involved in the process. These have been deeply described in \cite{Characteristics_time_scales}. Crucially, since the external metric can be explicitly given as a function of $m$ and $T$, the transition amplitude describing the tunneling process also depends on the same parameters. The surface $\Sigma$ in Figure \ref{fig:HR_spacetime} is formed by two flat 2-spheres joined at their boundary. We conventionally define these "upper" and "lower" boundary surfaces, associated with the future and past of the black hole, as $\Sigma_{+}$ and $\Sigma_{-}$.
\subsection{Discretization of $\Sigma$}
\label{subsec:Discretization_Sigma}
To write the transition amplitude explicitly and compute the crossing time, choosing a discretization of $\Sigma$ is necessary. We use the same discretization originally introduced in \cite{article:Christodoulou_Rovelli_Speziale_Vilensky_2016_planck_star_tunneling_time}. The geometry of the triangulation in terms of the Ashtekhar variables was completely derived in the same paper, to which we refer for further details. 

\medskip

\begin{figure}[h]
    \centering
        \begin{subfigure}[b]{6cm}
        \includegraphics[width=6cm]{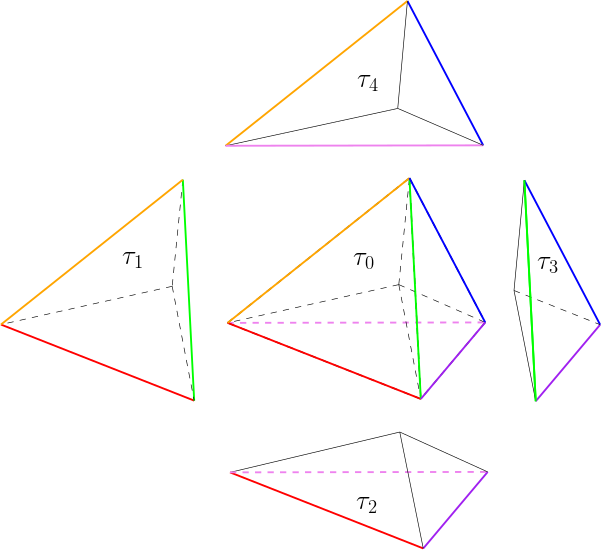}
    \end{subfigure}
   \caption{\label{fig:tetra_decomp}Regular tetrahedron $\tau_0$ split into four isosceles tetrahedra $\tau_1, \tau_2, \tau_3, \tau_4$. Both $\Sigma_{+}$ and $\Sigma_{-}$ are triangulated with such four isosceles tetrahedra.}
\end{figure}

\textbf{Triangulation}: each 2-sphere $\Sigma_{\pm}$ is first triangulated using a single equilateral flat tetrahedron. Then, the triangulation is refined by splitting each tetrahedron into four equal isosceles tetrahedra, as shown in Figure \ref{fig:tetra_decomp}. 
Therefore, the total surface $\Sigma$ is triangulated with eight boundary tetrahedra. The geometry is therefore composed of two 4-simplices (each one with zero 4-volume) joined by a tetrahedron. 

\medskip

\textbf{Two-complex}: the corresponding two-complex has two vertices contracted over a bulk intertwiner. The boundary graph is constituted of 16 links. Of these, 4 "angular" links $l_a$ connect the nodes between different vertices, while for each vertex, there are 6 "radial" links $l_{ab}^{\pm}$. So there are two different types of links. This discretization completely defines the spinfoam associated with the transition amplitude, which is described in section \ref{sec:Transition_amplitude}.
\subsection{Extrinsic boundary states}
\label{subsec:extr_bound_states}
After defining a discretization of $\Sigma$, it is possible to write down a coherent state describing the geometry. Among the possible definitions of coherent states, in \cite{article:Christodoulou_Rovelli_Speziale_Vilensky_2016_planck_star_tunneling_time}, the authors considered the "extrinsic" coherent states \cite{book:Rovelli_Vidotto_CLQG}, originally introduced by Thiemann \cite{Thiemann:2000ca}, parametrized as in \cite{Bianchi:2009ky} in terms of twisted geometries \cite{Freidel:2010aq}. These states depend on two unit-length source and target vectors $\vec{n}_s$, $\vec{n}_t$ and on a complex number $z$, which we write as:
\begin{equation}
\label{eq:z_def}
z = \eta + i \left( \beta + \gamma \zeta \right) \ . 
\end{equation}
In \eqref{eq:z_def}, $\eta \in \mathbb{R}^+$ is the dimensionless area of the triangular face dual to the link, $\zeta \in [0, 4 \pi)$ is the boost angle between the normals of the tetrahedra \cite{Dittrich:2008va, Rovelli:2010km} and $\beta$ is an extra rotation. We refer to \cite{article:Christodoulou_Rovelli_Speziale_Vilensky_2016_planck_star_tunneling_time} for the connection between the boost angle and the discretized holonomy along each link.  Extending analytically the definition of the Wigner matrices $D^j \left( h \right) $ to complex parameters, where $h \in SU(2)$, the extrinsic coherent states can be written as: 
\begin{align}
\label{eq:extrinsic_coh_state_original_def}
\Psi_{\sigma, n_s, n_t, z} \left( h \right)  = & \sum_{j} (2j+1) e^{- j(j+1)/2 \sigma} \times \nn \\
& {\rm Tr}[D^j(h)D^j(n_t e^{z\frac{\sigma_3}2} n_s^{-1})] \ ,
\end{align}
where $j$ is the spin attached to the link. The $SU(2)$ elements $n_s$, $n_t$ in \eqref{eq:extrinsic_coh_state_original_def} rotate the unit vector along the $\hat{z}$ axis into the source vector $\vec{n}_s$ and the target vector $\vec{n}_t$, respectively. The state on the graph is defined as the product of a factor \eqref{eq:extrinsic_coh_state_original_def} for each link.
%
%
%
Notice that apart from a phase factor, the ratio between two terms which differ by one unit in the component of the magnetic moment in the diagonal Wigner matrix $D^{j} \left( e^{z \frac{\sigma_3}{2}} \right)$ is:
\begin{equation}
\label{eq:condition_decoupling}
\frac{e^{\eta n}}{e^{\eta (n-1)}} = e^{\eta} > 10^3 \hspace{4mm} \text{for} \hspace{4mm} \eta \geq 7 \ ,
\end{equation}
where $n \in [-j, j]$. Therefore, when the real part $\eta$ of $\zeta$ is large enough, the trace in \eqref{eq:extrinsic_coh_state_original_def} is completely dominated by the highest magnetic moment component. As a consequence, the Wigner matrix can be approximated as:
\begin{equation}
\label{eq:condition_decoupling_Wigner}
D^{j}_{k,q} \left( e^{z \frac{\sigma_3}{2}} \right) \approx \delta_{k}^{j} \delta_{q}^{j} e^{z j} \ .
\end{equation}
When condition \eqref{eq:condition_decoupling_Wigner} is satisfied, the state \eqref{eq:extrinsic_coh_state_original_def} can be expressed as:
\begin{align}
\label{eq:extrinsic_coh_state_approx}
\Psi_{\sigma, n_s, n_t, z} \left( h \right)  \approx & \sum_{j} (2j+1) e^{- j(j+1)/2\sigma + z j} \times \nn \\
& \sum\limits_{n,m} D^{j}_{n,j}(n_t)   D^{j}_{m,n}(h) D^{j}_{j,m}(n_s^{-1})  \ ,
\end{align}
and the sum over $j$ in \eqref{eq:extrinsic_coh_state_approx} is peaked on the minimum of $j(j+1)/(2 \sigma)-\eta j$, which is
\begin{equation}
\label{eq:j_peaked_minimum}
j_m = \eta\sigma - \frac{1}{2} \ .
\end{equation}
\subsection{Balancing the spread}
\label{subsec:balancing_spread}
The quantity $\sigma$ in the definition of the extrinsic coherent states \eqref{eq:extrinsic_coh_state_original_def} plays an important role, as it determines whether the state is peaked on the area or the extrinsic curvature. We first consider a generic dependence:
\begin{equation}
\label{eq:alpha_trick_on_sigma}
\sigma = j_m^{-\alpha} \ ,
\end{equation}
where $\alpha \in \mathbb{R}$. From \eqref{eq:j_peaked_minimum} we find: 
\begin{equation}
\label{eq:relation_eta_jm}
\eta = j_m^{\alpha + 1} + \frac{j_m^{\alpha}}{2} \ .
\end{equation}
Recalling the relation existing between the spin and the area operator in LQG, it can be shown \cite{article:Bianchi_magliaro_perini_2010_coherent_spinnetwork} that the (relative) spread in the non-commuting areas and embedding data is:
\begin{equation}
\label{eq:dispersion_operators}
\frac{\Delta \zeta}{\langle \zeta \rangle} \sim j_m^{\frac{\alpha}{2}} \; ,\quad \frac{ \Delta A }{\langle A \rangle} \sim j_m^{- \left( \frac{\alpha}{2} + 1 \right)}  \ .
\end{equation}
The product between the two relative spreads \eqref{eq:dispersion_operators} is independent of $\alpha$ and goes to zero in the large $j_m$ limit. A good requirement to recover the semiclassical behavior is that the relative dispersions \eqref{eq:dispersion_operators} vanish for $j_m \longrightarrow \infty$. Along with condition \eqref{eq:alpha_trick_on_sigma} and requiring that both $\eta$ and $j_m$ are large, this results in the range $\alpha \in [-1, 0]$ for the extrinsic coherent states \eqref{eq:extrinsic_coh_state_original_def} to behave semiclassically. In \cite{article:Christodoulou_Rovelli_Speziale_Vilensky_2016_planck_star_tunneling_time}, the choice made was $\alpha = - \frac{1}{2}$ to peak the state both on the area and on the extrinsic curvature in the large $j_m$ limit. Crucially, notice that with this choice, when $j_m$ is small, condition \eqref{eq:condition_decoupling_Wigner} is not a valid approximation. Therefore, we consider the case $\alpha > 0$ so that \eqref{eq:condition_decoupling_Wigner} is valid. From \eqref{eq:dispersion_operators}, this choice leads to a sharp area operator and an extrinsic spread curvature. In the language of \cite{article:Bianchi_magliaro_perini_2010_coherent_spinnetwork}, this corresponds to a large heat-kernel time. Physically, this means increasing the quantum spread associated with the boost angle operator between $\Sigma_{-}$ and $\Sigma_{+}$. From \eqref{eq:alpha_trick_on_sigma} and \eqref{eq:relation_eta_jm} we obtain:
\begin{equation}
\frac{-j(j+1)}{2 \sigma} + j\,\eta = - \frac{j_m^{\alpha}}{2} \big( j - j_m \big)^2 \, + \frac{j_m^{\alpha + 2}}{2} \ ,
\label{eq:quadratic}
\end{equation}
where the last term is absorbed in the normalization factor of the amplitude. In the transition amplitude, every link of the boundary graph has an extrinsic coherent state \eqref{eq:extrinsic_coh_state_original_def} associated with it. Therefore, each boundary link has an independent (infinite) sum. The condition $\alpha > 0$ increases the factor $j_m^{\alpha}$, so putting a small half-integer cut-off $K$ around $j_m$ allows a good amplitude approximation. 
This corresponds to the physical limit in which we recover the intrinsic coherent states from the extrinsic ones.

\medskip

The semiclassical condition $\alpha \in [-1,0]$ implies that states \eqref{eq:extrinsic_coh_state_original_def} become rapidly spread around $j_m$ for large $j_m$. In such a regime, a possible approach to computing the amplitude numerically could consist in using importance sampling Monte Carlo to overcome the multiple independent sums over boundary links. However, one has to deal with the sign problem in such a case. Alternatively, one could use the analytical methods developed in \cite{Fabio_PhD_thesis, christodoulou2023geometry} suitable for the semiclassical regime.
\subsection{Normal orientation}
\label{subsec:normal_orientation}
\begin{figure}[h]
\centering
\begin{subfigure}[b]{6cm}
\includegraphics[width=6cm]{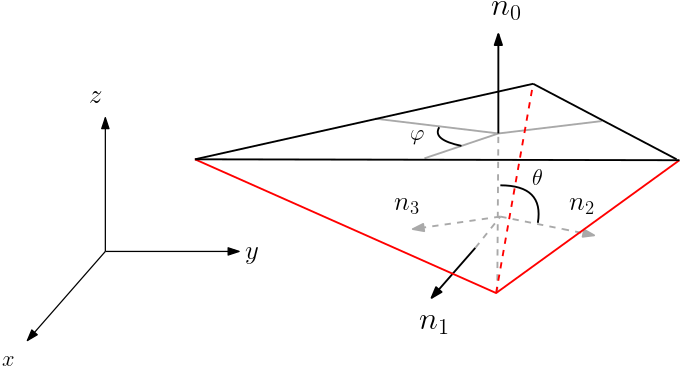}
\end{subfigure}   
\caption{\label{fig:tetra_orient}Normals orientation in each isosceles tetrahedron.}
\end{figure}

In this Section, we define the orientation of normals to the boundary tetrahedra in the triangulation described in Section \ref{subsec:Discretization_Sigma}. These were originally computed in \cite{article:Christodoulou_Rovelli_Speziale_Vilensky_2016_planck_star_tunneling_time}. We parametrize the Wigner matrix as in \cite{book:varshalovic}:
\begin{equation}
\label{eq:wigner_matrix}
D^{j}_{m,j}(n) = D^{j}_{m,j}(\phi, \theta,-\phi)= e^{-i m \phi} d^j_{m,j}(\theta) e^{i j \phi} \ ,
\end{equation} 
where $d^j_{m,j}(\theta)$ is the small Wigner matrix \cite{book:varshalovic, GraphMethods}. The $\vec{n}$ vector can be parametrized with the polar angles as usual:
$$
\vec{n} = (\sin\theta\cos\phi, \sin\theta\sin\phi,\cos\theta) \ .
$$
Using the orientation of Figure \ref{fig:tetra_orient}, after some calculation \cite{article:Christodoulou_Rovelli_Speziale_Vilensky_2016_planck_star_tunneling_time} the following values are obtained:
\begin{eqnarray}
\label{eq:normal_values}
\vec n_0&=&(0,0),  \\
\vec n_k&=&\left(\arccos{\scriptstyle
\left[-\sqrt{\frac23}\right]},\ \ \varphi_k\right),  
\end{eqnarray}
with $k=1,2,3$ and
\begin{equation}
\varphi_1=0, \ \ \ \ \ \varphi_2=\frac23 \pi, \ \ \ \ \   \varphi_3=-\frac23 \pi \ .
\end{equation}
The required value for $\beta$ in \eqref{eq:z_def} turns out to be $\beta = 0$ for the equilateral faces, and $\beta = \varphi_{k} - \varphi_{k'}$ for the isosceles faces. This extra rotation along the $\hat{z}$ axis must match the triangles in the $(x,y)$ plane. The effect of such rotation is such that we can replace the $SU(2)$ element $n$ in \eqref{eq:wigner_matrix} with another element $\tau$, whose third component is zero:
\begin{equation}
\label{eq:wigner_matrix_extra_rotation}
D^{j}_{m,j}(\tau) = D^{j}_{m,j}(\phi, \theta, 0)= e^{-i m \phi} d^j_{m,j}(\theta) \ .
\end{equation} 
For the target on the same link, we add a parity transformation so that $\theta \longrightarrow  \theta - \frac{\pi}{2}$. Finally, we define the Livine-Speziale coherent intertwiner coefficient in the recoupling channel $i$ with all outgoing links as:
\begin{align}
\label{eq:coherent_state_coefficient}
\psi_{i}(\tau) & = \sum_{m_a} \Wfour{j_1}{j_2}{j_3}{j_4}{m_1}{m_2}{m_3}{m_4}{i} \prod_{a=1}^4
D^{j_a}_{m_a,j_a}(\tau) \nn \\
& =  \raisebox{-5mm}{\includegraphics[width=2.8cm]{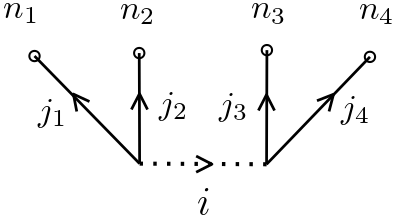}} \ ,
\end{align}
where $D^{j}_{m,j}(\tau)$ has been defined in \eqref{eq:wigner_matrix_extra_rotation}. The coefficient \eqref{eq:coherent_state_coefficient} encodes the orientation of the normals in the final amplitude. The definition of the 4jm Wigner symbol is reported in Appendix \ref{app:SU(2)_symbols}. 
\section{Transition amplitude}
\label{sec:Transition_amplitude}
\subsection{The EPRL vertex amplitude}
\label{subsec:EPRL_vertex_amplitude}
We write the EPRL vertex amplitude using the graphical notation discussed in detail in \cite{Review_numerical_LQG}:
\begin{align}
\label{eq:EPRL_vertex_amplitude}
& 
V_{\gamma} \left(j_f, \, i_e ; \Delta l \right) = \\
& = 
\sum_{\substack{j_f\leq l_{f}\leq j_f + \Delta l \\ k_a}} \left( \prod_{a=1}^{4} d_{k_{a}} B_{4}(j_{f}, l_{f}; i_{a}, k_{a}; \gamma) \right)  \{15j\}_{l_{f}, k_{a}} \nn \\ \nn
& =   
\sum_{j_f\leq l_{f}\leq j_f + \Delta l} 
\raisebox{-2cm}{\includegraphics[width=5cm]{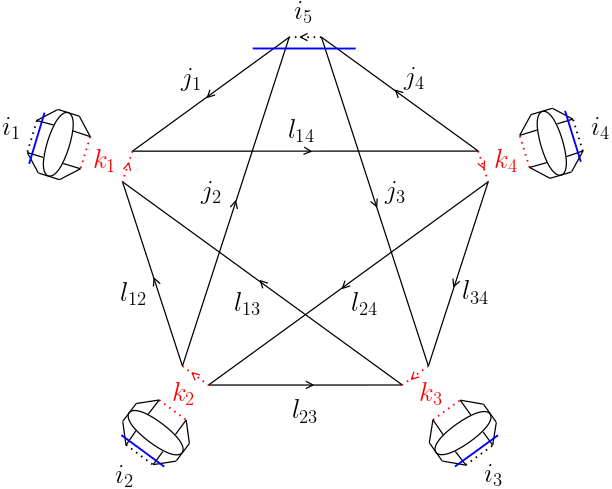}} \ ,
\end{align}
where $a = 1 \dots 4$, $e = 1 \dots 5$, $q = 2 \dots 5$. The dependence of the amplitude on the Barbero-Immirzi parameter $\gamma$ has been indicated using a label. The definition of the 15j Wigner symbol is reported in Appendix \ref{app:SU(2)_symbols}, while the $B_4$ function is defined in Appendix \ref{app:booster}. We compute the EPRL vertices \eqref{eq:EPRL_vertex_amplitude} with the numerical framework \texttt{sl2cfoam-next} \cite{Francesco_draft_new_code}. We define the coherent amplitude as the vertex amplitude \eqref{eq:EPRL_vertex_amplitude} contracted with coherent states coefficients \eqref{eq:coherent_state_coefficient} over all nodes except one\footnote{The contraction over all nodes of the vertex is usually considered, but for the present context the definition \eqref{eq:coherent_vertex_amplitude} is more convenient}:
\begin{align}
\label{eq:coherent_vertex_amplitude}
V^{coh}_{\gamma, n_f} \left(j_f, i_5; \Delta l \right) = \sum_{i_a} V \left( \prod_{a=1}^{4} d_{i_a} \psi_{i_a} (\tau_f) \right) \ .
\end{align}
The graphical notation of the coherent amplitude \eqref{eq:coherent_vertex_amplitude} is easily obtained from \eqref{eq:EPRL_vertex_amplitude} and \eqref{eq:coherent_state_coefficient}. We do not report it explicitly for the single vertex amplitude. Instead, we use it directly in the Black-to-White hole transition amplitude described in Section \ref{subsec:BW_hole_transition_amplitude}.
\subsection{The Black-to-White hole transition amplitude}
\label{subsec:BW_hole_transition_amplitude}
The Black-to-White hole transition amplitude is obtained by contracting the amplitude associated with the spinfoam described in Section \ref{subsec:Discretization_Sigma} with the coherent boundary state described in Section \ref{subsec:extr_bound_states}, according to the usual procedure in covariant LQG \cite{book:Rovelli_Vidotto_CLQG}. We refer to the original article \cite{article:Christodoulou_Rovelli_Speziale_Vilensky_2016_planck_star_tunneling_time} for a description of all the necessary steps.

\medskip

With the definition of the EPRL vertex amplitude \eqref{eq:EPRL_vertex_amplitude}, it is possible to write the Black-to-White hole transition amplitude in a suitable form for a numerical evaluation, transforming the original 24j Wigner symbol into the contraction of two (linear superposition of) 15j symbols over a bulk intertwiner. Aside from the normals to the boundary tetrahedra, each link has an associated spin $j$ and boost angle $\zeta$. In \cite{article:Christodoulou_Rovelli_Speziale_Vilensky_2016_planck_star_tunneling_time}, it was shown that when condition \eqref{eq:condition_decoupling_Wigner} is satisfied, the dependence on the parameters $m, T$ is decoupled from the combinatorial structure of the two-complex. This greatly simplifies the numerical evaluation and leads to a factorization of the amplitude in the form of (a spin-sum over) a "weight function" $w$ that multiplies the factor associated with the contraction of the two vertex amplitudes. We define the weight function as:
\begin{widetext}
\begin{align}
\label{eq:weight_factor}
w_{\alpha} (j_a, j^\pm_{ab}, j_{\pm}, j_{0}, \zeta_{\pm}, \zeta_0) = c_{\alpha} \left(j_{\pm}, j_0 \right)  \left( \prod_{a=1}^{4} d_{j_a} e^{- \frac{j_0^{\alpha}}{2} {(j_a - j_0)}^2} e^{i \gamma \zeta_0 j_a }\right) \left( \prod_{ab,\pm} d_{j_{ab}^\pm} e^{- \frac{j_{\pm}^{\alpha}}{2} (j_{ab}^\pm - j_{\pm})^2} e^{i \gamma \zeta_{\pm} j_{ab}^\pm}\right) \ ,
\end{align}
where $b = 2, 3, 4$, $a \neq b$. The data $(j_{0}, \zeta_0)$ label the 4 angular links, while $(j_{\pm}, \zeta_{\pm})$ are associated with the radial links. The normalization factor inherited from the boundary state is:
\begin{equation}
\label{eq:normalization_factor_states}
c_{\alpha} \left(j_{\pm}, j_0 \right) = \left( e^{\frac{j_0^{\alpha + 2}}{2}} \right)^4 \left( e^{ \frac{j_{\pm}^{\alpha + 2}}{2}} \right)^{12} \ ,
\end{equation}
which corresponds to the last factor in \eqref{eq:quadratic} inherited by each link. We write the Black-to-White hole transition amplitude as:
\begin{align}
\label{eq:BW_amplitude}
&
W_{\alpha} (j_{\pm}, j_{0}, \zeta_{\pm}, \zeta_0; \Delta l) = \sum_{ j_{ab}^{\pm}, j_{a} } w_{\alpha} \left( \sum_{i_5} d_{i_5} \prod_{\pm} V_{\gamma,  n_{f}^{\pm}}^{coh} \left(j_{ab}^{\pm}, j_{a}, i_5; \Delta l \right) \right) \nn \\
& = \sum_{ j_{ab}^{\pm}, j_{a} } w_{\alpha} \sum_{j_{ab}^{\pm}\leq l_{ab}^{\pm}\leq j_{ab}^{\pm} + \Delta l} \hspace{6mm}
\raisebox{-3.6cm}{\includegraphics[width=6cm]{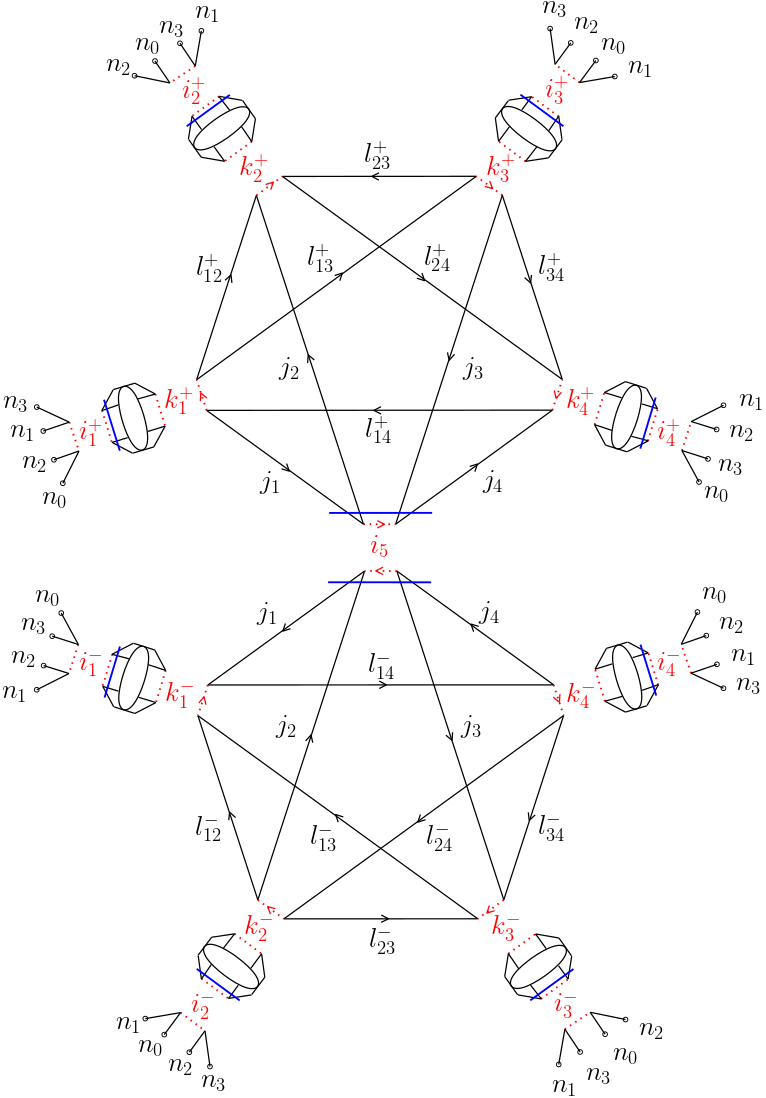}} \ .
\end{align}
\end{widetext}
The graphical notation in \eqref{eq:BW_amplitude} emphasizes how the normals to the tetrahedra discussed in Section \ref{subsec:normal_orientation} are associated with the faces of the triangulation discussed in Section \ref{sec:Geometry}. The intertwiner $i_5$ is dual to tetrahedron $\tau_0$ in Figure \ref{fig:tetra_decomp}. In \cite{Frisus_BW_repo}, we provide a Mathematica notebook to re-construct the full geometry (we thank Pietro Dona for the help with the notebook). Finally, we emphasize that the compact notation used in the first line of \eqref{eq:BW_amplitude} does not specify if each link is a source or a target, but this is clarified with the graphical notation. 

\medskip

As mentioned in Section \ref{sec:Geometry}, the external geometry is entirely defined by the parameters $m, T$. Therefore, these two parameters entirely determine the amplitude \eqref{eq:BW_amplitude}. The relationship between these and the variables $j_0, j_{\pm}, \gamma, \zeta_{\pm}, \zeta_0$ is provided by the following relations:
\begin{align}
\label{eq:relations_m_T_spins_j0}
j_0 & = \frac{m^2 \left( 1 + e^{- \frac{T}{2m}} \right)^2}{2 \gamma}  \ ,   \\
\label{eq:relations_m_T_spins_jpm}
j_{\pm} & = \frac{j_0}{\sqrt{6}} \ , \\ 
\label{eq:relations_m_T_spins_zeta0}
\zeta_0 & = \frac{T}{2m} \ , \\
\label{eq:relations_m_T_spins_zetapm}
\zeta_{\pm} & = \mp \frac{32}{9} \sqrt{6} \ .
\end{align}
Despite the equality sign, in equation \eqref{eq:relations_m_T_spins_j0} it was used the well-known approximation $ A =  8 \pi \gamma \sqrt{j \left( j + 1 \right)} \approx 8 \pi \gamma j $. A few comments are in order.

\medskip

The discrete nature of spin has interesting consequences. For example, relation \eqref{eq:relations_m_T_spins_jpm} implies that $j_{0}$ and $j_{\pm}$ cannot be both half-integer numbers. These are the terms on which the boundary coherent states 
\eqref{eq:extrinsic_coh_state_original_def} are peaked. Furthermore, triangular inequalities impose constraints on the allowed spins configurations.
\section{Computing the amplitude}
\label{sec:algorithm}
\subsection{The numerical algorithm}
The algorithm to calculate the black-to-white hole transition amplitude \eqref{eq:BW_amplitude} as a function of $T$ can be divided into three main steps. The strategy is similar to the one outlined in \cite{Review_numerical_LQG}. The core idea is to separate the computation of the EPRL vertex amplitudes \eqref{eq:EPRL_vertex_amplitude} from the contraction along the intertwiners of each vertex, which is typically much less resource-demanding than the former. The code used for all calculations in this paper is public and available on \texttt{GitHub} \cite{Frisus_BW_repo}. 

\medskip

\begin{figure}[h]
\centering
\begin{subfigure}[b]{.49\textwidth}
\includegraphics[width=\linewidth]{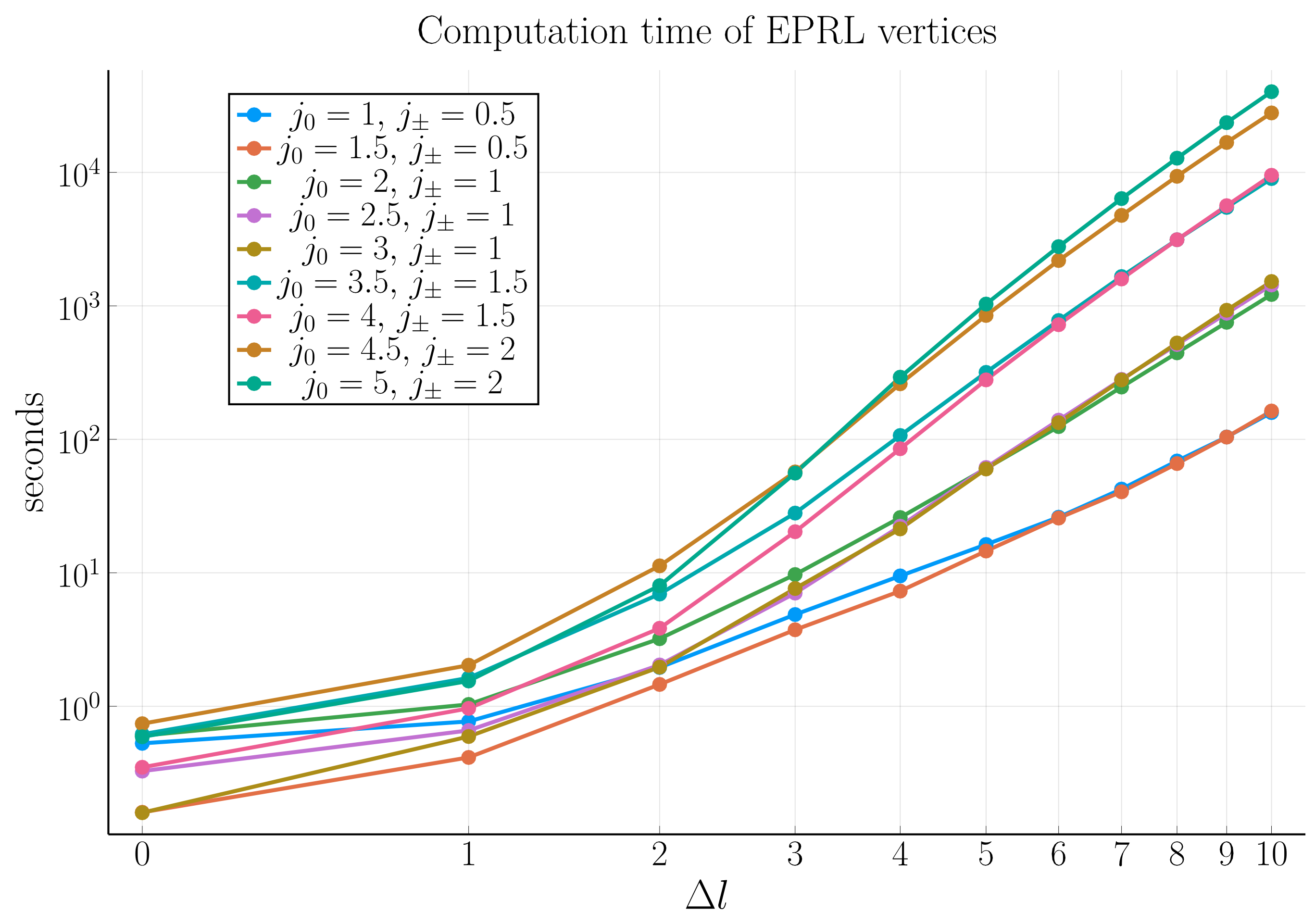}
\end{subfigure} 
\begin{subfigure}[b]{.49\textwidth}
\includegraphics[width=\linewidth]{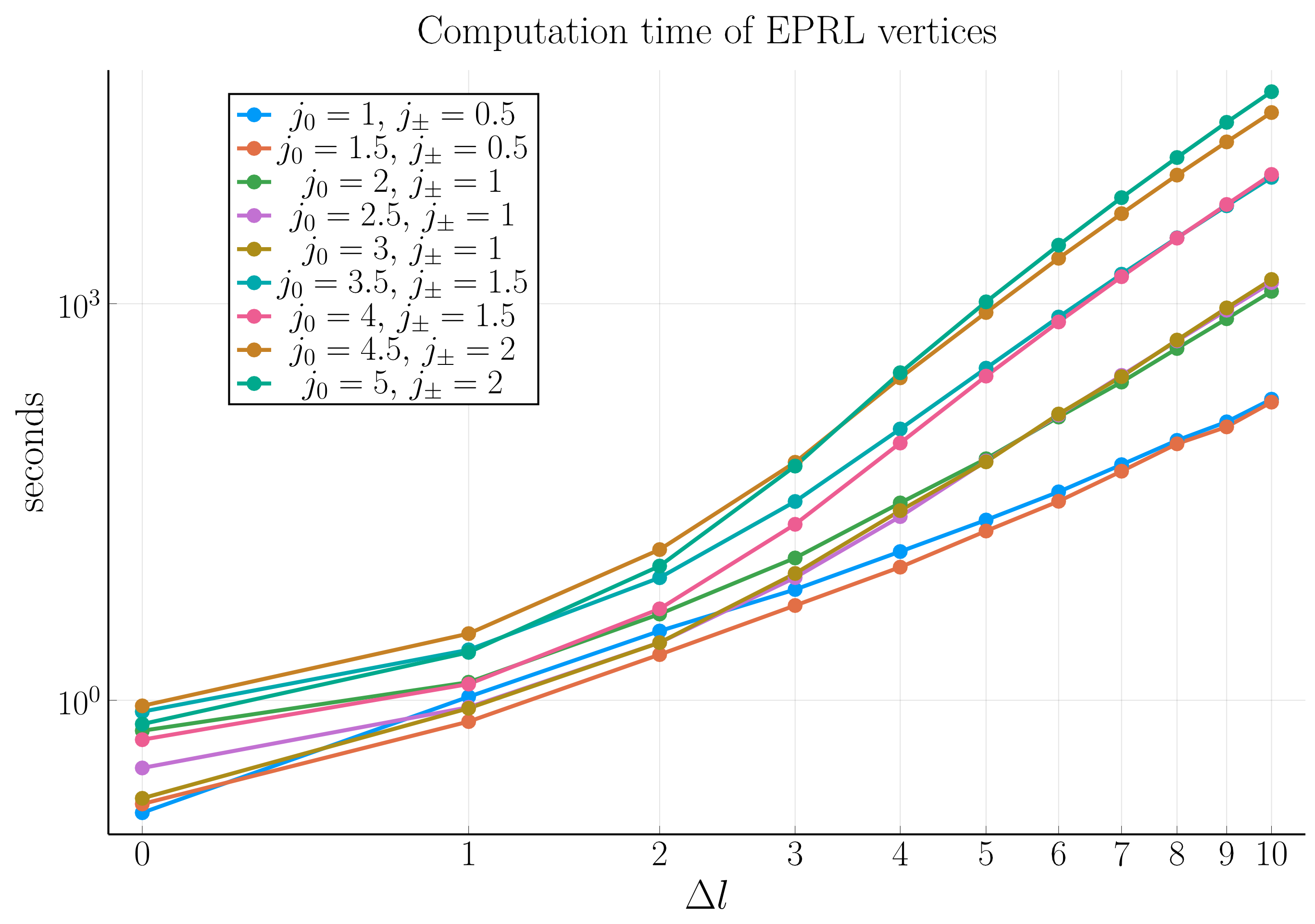}
\end{subfigure} 
\caption{\label{fig:comp_times}Log-log plot of computational time required for algorithm \ref{alg_part_1}. \textbf{Top:} case $\gamma = 1$. \textbf{Bottom:} case $\gamma = 5$.}
\end{figure}

The first step is to pre-calculate all the necessary EPRL vertex tensors \eqref{eq:EPRL_vertex_amplitude}. With the term "tensor," we refer to the multidimensional array consisting of the vertex amplitude computed for all the possible values of intertwiners. The flowchart is reported in \ref{alg_part_1}. 
\begin{algorithm}[H]
\caption{Part 1: computing the EPRL vertices}
\label{alg_part_1}
\begin{algorithmic}[1]
\State At fixed Immirzi constant $\gamma$, choose the parameters $\alpha$, $K_{0}$, $K_{\pm}$, $j_{0}^{min}$, $j_{0}^{max}$ as described in Section \ref{subsec:balancing_spread}
\State Set a maximum value $\Delta l^{max}$
\For{$j_0 \in \{ j_{0}^{min} , j_{0}^{min} + \frac{1}{2}, \dots j_{0}^{max} \}$}
\State Calculate $j_{\pm}$ from \eqref{eq:relations_m_T_spins_jpm} and round to the nearest half-integer 
\For{$\Delta l \in \{ 0, 1 \dots \Delta l^{max} \}$}
\State Compute all the EPRL vertex amplitudes \eqref{eq:EPRL_vertex_amplitude} with $j_{a} \in \left[ j_{0} - K_{0}, \ j_{0} + K_{0} \right]$, $j_{ab} \in \left[ j_{\pm}-K_{\pm}, \ j_{\pm} + K_{\pm} \right]$
\State Dump the vertices to disk
\EndFor 
\State \textbf{end}
\EndFor 
\State \textbf{end}
\end{algorithmic}
\end{algorithm}
This is the most demanding step regarding computational resources and time complexity. The calculation of the vertex tensors has been performed with the \texttt{sl2cfoam-next} library \cite{Francesco_draft_new_code} on the \textit{Cedar}, \textit{Graham} and \textit{Narval} Compute Canada superclusters. We employed a hybrid parallelization scheme, distributing the workload on multiple processes and eventually exploiting various threads for each task. In \cite{Frisus_BW_repo}, we also provide a code that automatically distributes the calculation of the vertex tensors to multiple machines. In Figure \ref{fig:comp_times} we report the computational time of algorithm \ref{alg_part_1} for $j_0^{min} = 1$, $j_0^{max} = 5$, $\Delta l^{max} = 10$, $K_0 = 0.5$, $K_{\pm}=0.5$. Each curve represents the seconds required to compute all the vertex amplitudes \eqref{eq:EPRL_vertex_amplitude} centered around different spins configurations $j_0$, $ j_{\pm}$. In the top panel, we show the results for $\gamma = 1$, while in the bottom one, we report the time for $\gamma = 5$. The two cases have approximately the same trend, even if the calculation for larger values of the Barbero-Immirzi parameter requires more time. The results in Figure \ref{fig:comp_times} were estimated by distributing the computation of the vertices over $64$ CPUs AMD Rome 7532 @ 2.40 GHz 256M cache L3.

\medskip

The second step consists in contracting the stored vertex tensors. The contraction is performed between the vertices and the coherent state coefficients \eqref{eq:coherent_state_coefficient} according to the spinfoam structure described in Section \ref{sec:Geometry}. This is illustrated in the flowchart \ref{alg_part_2}.
\begin{algorithm}[H]
\caption{Part 2: contracting the EPRL vertices}
\label{alg_part_2}
\begin{algorithmic}[1]
\For{the spins configurations considered in algorithm \ref{alg_part_1}}
\State Compute all the coherent state coefficients \eqref{eq:coherent_state_coefficient} with $j_{a} \in \left[ j_{0} - K_{0}, \ j_{0} + K_{0} \right]$, $j_{ab} \in \left[ j_{\pm}-K_{\pm}, \ j_{\pm} + K_{\pm} \right]$
\For{$\Delta l \in \{ 0, 1 \dots \Delta l^{max} \}$}
\State Retrieve the amplitudes stored during algorithm \ref{alg_part_1} and load them into memory
\State Contract the vertices with the coherent states to obtain two coherent amplitudes \eqref{eq:coherent_vertex_amplitude} as in \eqref{eq:BW_amplitude}
\State Contract the coherent amplitudes along $i_5$
\State Dump the results to disk
\EndFor 
\State \textbf{end}
\EndFor 
\State \textbf{end}
\end{algorithmic}
\end{algorithm}
The final result of this step is a set of complex numbers, which correspond to the term in round brackets in \eqref{eq:BW_amplitude} multiplying the weight factor $w$. This represents the pure "spinfoam contribution" to the amplitude, which is contracted with the weight factor of the boundary states. 

\medskip

\begin{figure*}[!hbtp]
    \centering
    \begin{subfigure}[b]{.49\textwidth}
        \includegraphics[width=\linewidth]{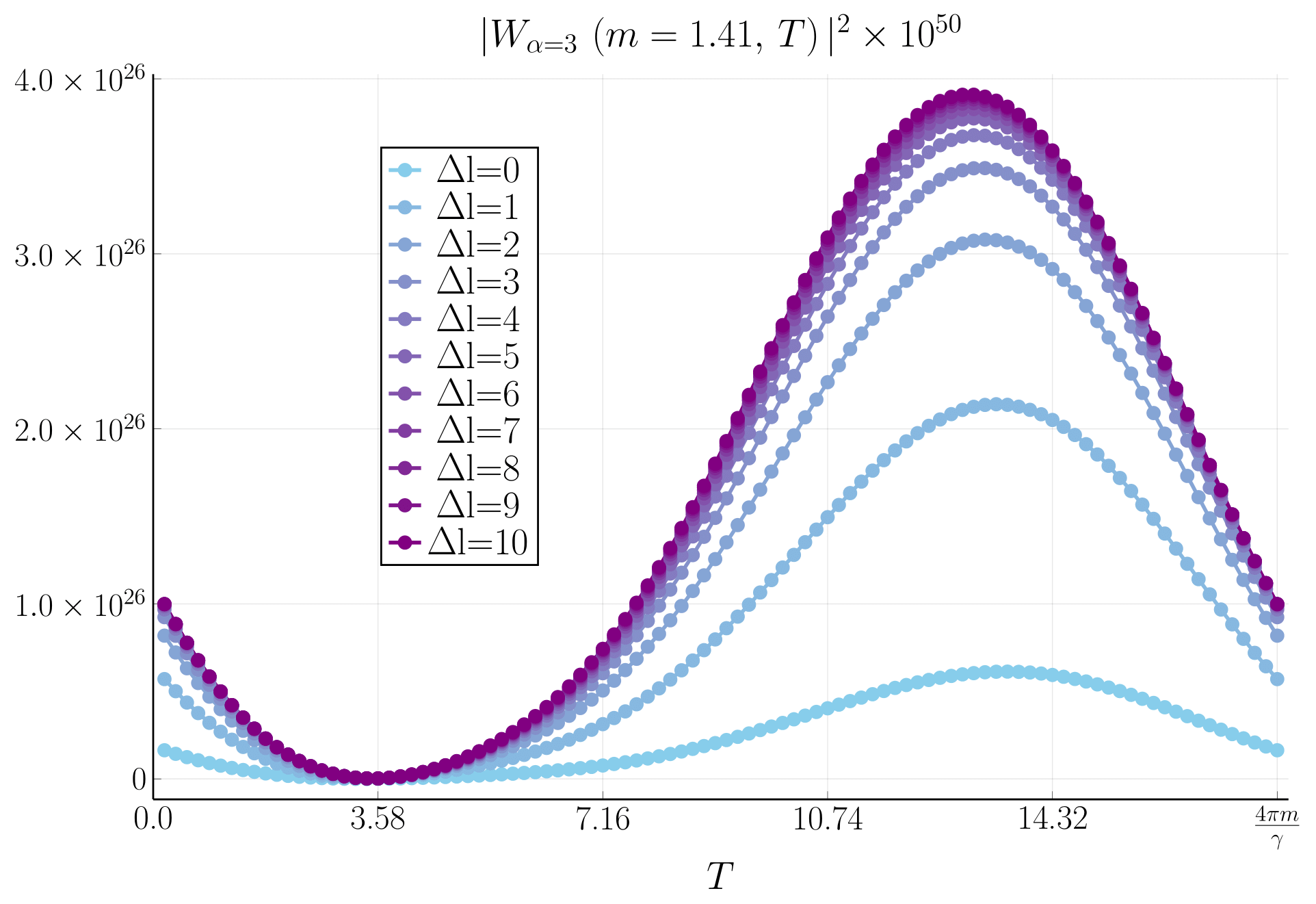}
    \end{subfigure}  
    \begin{subfigure}[b]{.49\textwidth}
        \includegraphics[width=\linewidth]{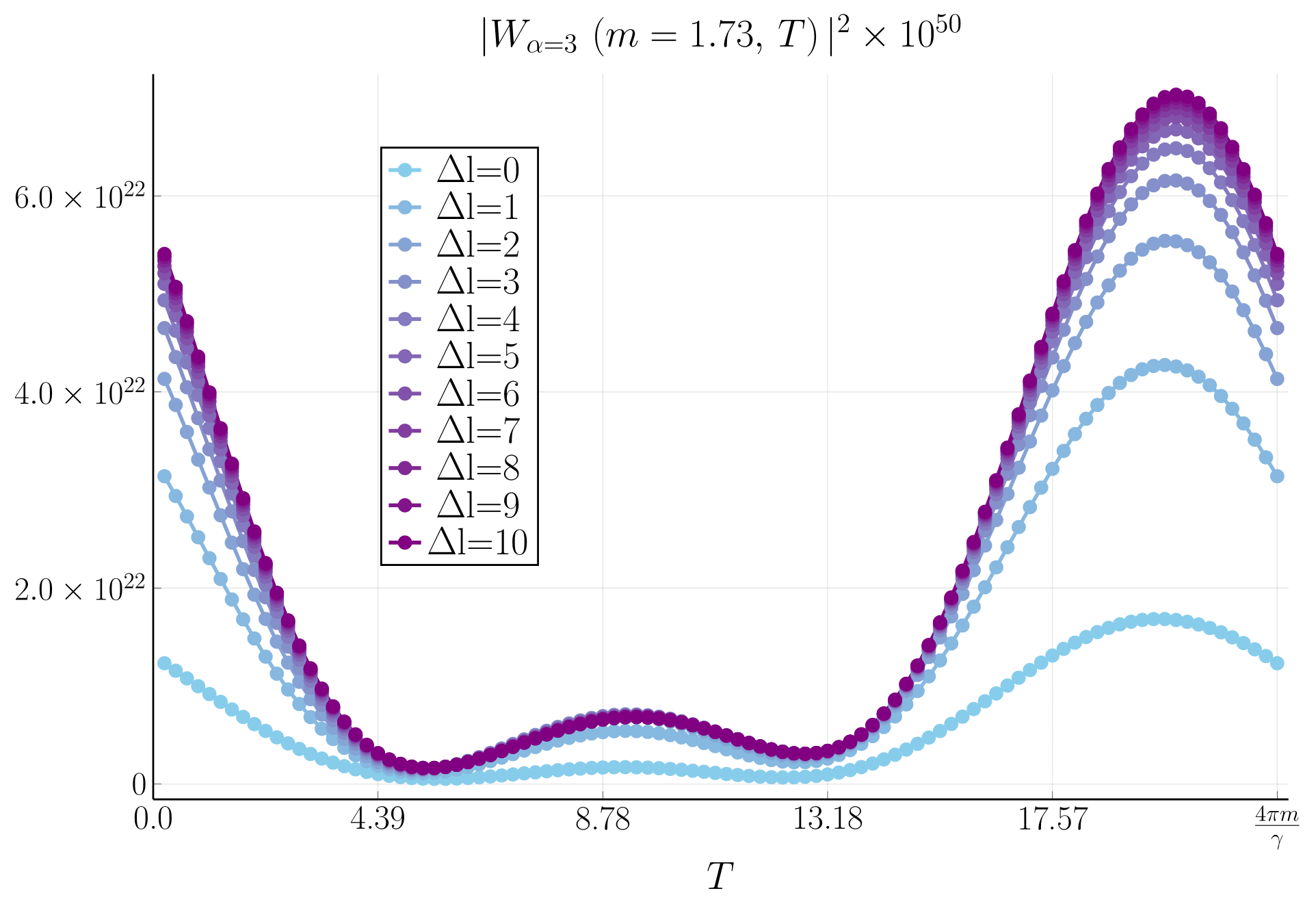}
    \end{subfigure}  
    \begin{subfigure}[b]{.49\textwidth}
        \includegraphics[width=\linewidth]{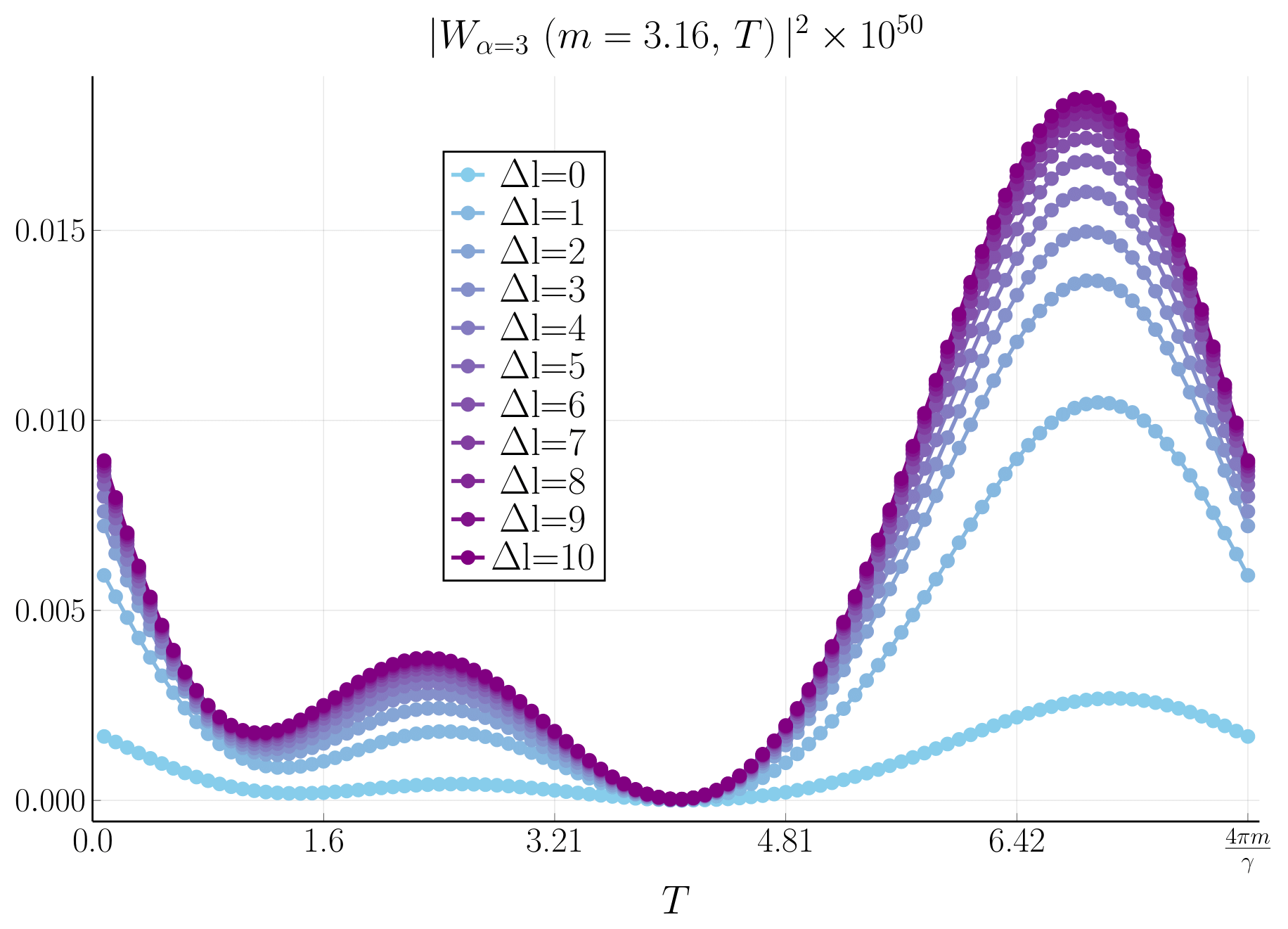}
    \end{subfigure}  
    \begin{subfigure}[b]{.49\textwidth}
        \includegraphics[width=\linewidth]{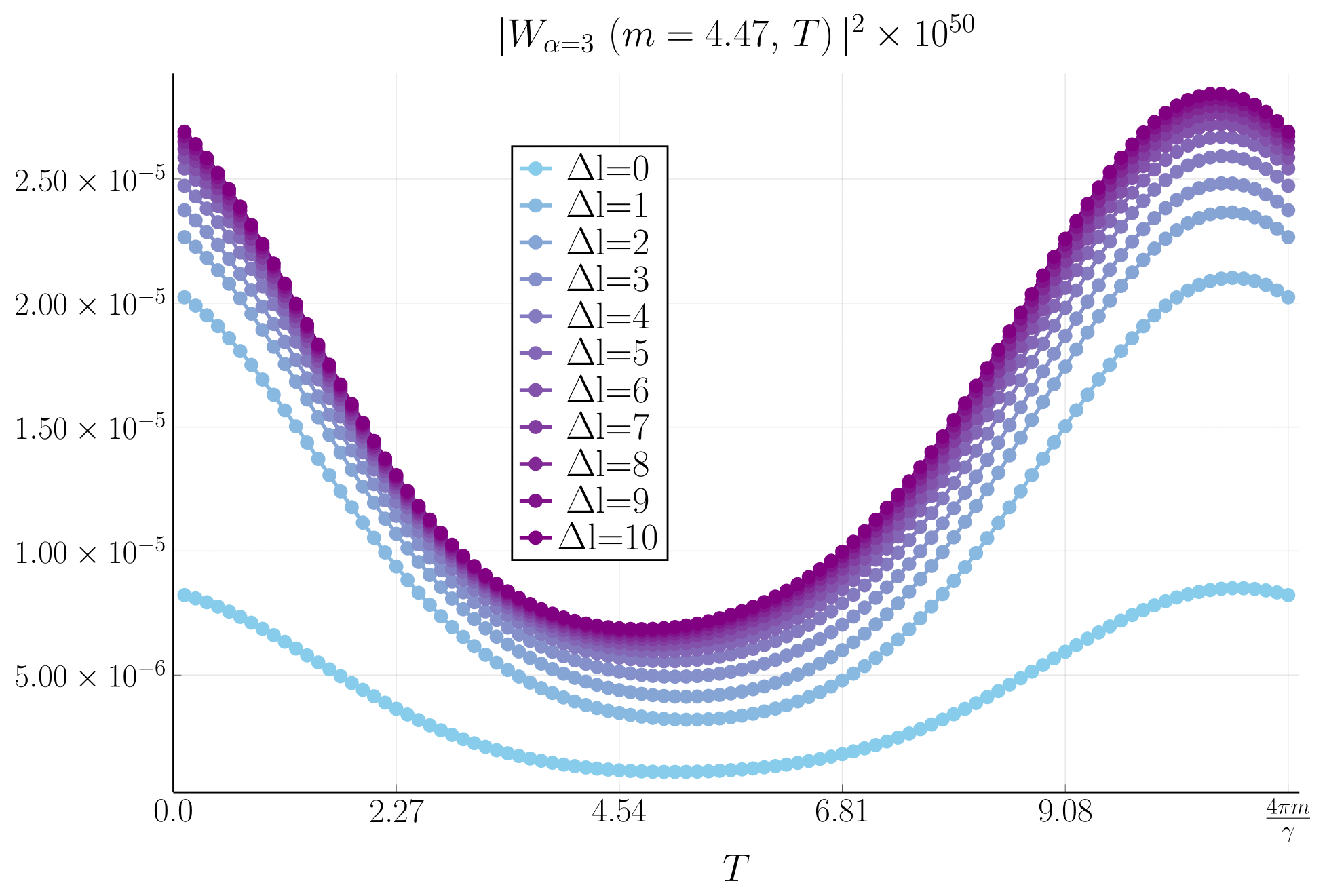}
    \end{subfigure}      
    \caption{\label{fig:W_squared} Black-to-White hole transition amplitude \eqref{eq:BW_amplitude} computed with the algorithm described in Section \ref{sec:algorithm}. In the partition \eqref{eq:W_partition}, it was used $N=100$. The truncation parameter's value $\Delta l=10$ reasonably approximates the amplitude. \textbf{Top:} case $\gamma = 1$. \textbf{Bottom:} case $\gamma = 5$.}
\end{figure*}

The third and final step consists in computing the weight factor \eqref{eq:weight_factor} and assembling the amplitude by retrieving all the pieces previously computed. As noticed in \cite{article:Christodoulou_Rovelli_Speziale_Vilensky_2016_planck_star_tunneling_time}, the amplitude \eqref{eq:BW_amplitude} is periodic in the extrinsic curvature angle $\zeta_0$ with period $\frac{4 \pi m}{\gamma}$ in $T$. This is a consequence of the discretization discussed in Section \ref{subsec:Discretization_Sigma}. Following the strategy of \cite{Characteristics_time_scales, christodoulou2023geometry, Fabio_PhD_thesis}, we restrict the validity of the amplitude \eqref{eq:BW_amplitude} to a single period over $T$. We choose a parameter $N >> 1$ and divide the interval $\left[ 0, \frac{4 \pi m}{\gamma} \right]$ into $N$ equal sub-intervals with constant $T$:
\begin{equation}
\label{eq:T_partition}
0 \equiv T_0 < T_1 < T_2 \dots < T_N \equiv \frac{4 \pi m}{\gamma} \ , \ N >> 1 \ .
\end{equation}
The weight factor is computed in each sub-interval, and the amplitude is assembled by retrieving and assembling all pieces. The value of $m$ is computed using \eqref{eq:relations_m_T_spins_j0} disregarding the $T$ dependence, as the term $e^{- \frac{T}{2m}}$ rapidly becomes negligible as a function of $T$. As shown in \ref{subsec:crossing_time_large_j}, neglecting this term results in a shift of the crossing time, but it does not alter the functional dependence on $T$. The flowchart is shown in \ref{alg_part_3}.
\begin{algorithm}[H]
\caption{Part 3: Assembling the B-W amplitude}
\label{alg_part_3}
\begin{algorithmic}[1]
\For{the spins configurations considered in algorithm \ref{alg_part_1}}
\State Compute $m$ using \eqref{eq:relations_m_T_spins_j0} 
\State Choose a parameter $N >> 1$ and divide the first period in $T$ according to \eqref{eq:T_partition}
\For{each sub-interval} 
\State Compute and store the weight factor \eqref{eq:weight_factor}
\State Assemble the amplitude \eqref{eq:BW_amplitude} retrieving the data stored in algorithm \ref{alg_part_2}
\EndFor
\State \textbf{end}
\State Dump the amplitudes to disk
\EndFor
\State \textbf{end}
\end{algorithmic}
\end{algorithm}
The final result of algorithm \ref{alg_part_3} is a set of amplitudes:
\begin{equation}
\label{eq:W_partition}
\{ W_{T_i} \}_{i=0}^{N} \equiv \{ W_{T_0},  W_{T_1} \dots  W_{T_N} \} \ , \ N >> 1 
\end{equation}
corresponding to the partition \eqref{eq:T_partition}, which can be used to compute the physical observables depending on the amplitude 
\eqref{eq:BW_amplitude} as a function of $T$. We discuss one example in Section \ref{sec:crossing_time}. In figure \ref{fig:W_squared}, we display the results of the (rescaled) amplitude computed using the algorithm \ref{alg_part_3} and $N=100$ in the partition \eqref{eq:T_partition}. Each point corresponds to the squared absolute value of the elements in the partition \eqref{eq:W_partition}. Notice that in Figure \ref{fig:W_squared} the amplitude is rescaled so that the \texttt{julia} package \texttt{Plots.jl} displays the value correctly. A few comments are in order.

\medskip

The exact value of the amplitude is recovered in the limit $\Delta l \longrightarrow \infty$. This parameter is introduced in the EPRL vertex amplitude \eqref{eq:EPRL_vertex_amplitude} as a homogeneous truncation to approximate the unbounded convergent sums over the virtual spins $l_f$ \cite{Speziale2016}. The role of this parameter has been deeply discussed in many papers focusing on numerical computations of spinfoam amplitudes \cite{Review_numerical_LQG, Dona2018, Francesco_draft_new_code, Frisoni_2023, Self_energy_paper, Radiative_corrections_paper}. The amplitude becomes constant as $m$ increases since the quantum fluctuations are suppressed as the spin grows. That is, approaching the semiclassical limit as discussed in \ref{subsec:balancing_spread}. With the expression "quantum fluctuations," we refer to the terms in the sum \eqref{eq:extrinsic_coh_state_approx} defining the extrinsic coherent state with $j \neq j_{m}$. In \eqref{eq:BW_amplitude} we have such a sum for each link, where $j_m$ corresponds to $j_0$ in the case of angular links and $j_{\pm}$ for the radial ones.

\medskip

Higher orders in the vertex expansion are necessary to investigate larger $T$ values and resolve (at least partially) the periodicity of amplitude \eqref{eq:BW_amplitude} in $T$. An example of complete derivation of the black-to-white hole transition amplitude with a finer triangulation has been derived in \cite{Black_Hole_part_2}. Unfortunately, the level of complexity in the numerical evaluation of the amplitude grows very quickly as the triangulation refinement increases.
\section{The crossing time}
\label{sec:crossing_time}
In this Section, we estimate the crossing time both numerically and analytically. It represents the characteristic time scale for the transition when it takes place. For an accurate and comprehensive physical description of this observable (as well as other time scales involved in the tunneling process), we refer to \cite{Characteristics_time_scales}. In the following, we do not explicitly indicate the dependence on the truncation parameter $\Delta l$, implying that the latter has been fixed to $\Delta l = 10$ providing a reasonable amplitude estimate. According to the probabilistic interpretation of the transition amplitude developed by Oeckl \cite{Oeckl_2003, Oeckl_2008}, explicitly applied to the black-to-white hole transition in \cite{Oeckl_2018}, we first define a conditional probability distribution:
\begin{equation}
\label{eq:cond_prob_dist}
P_{\alpha} \left( m | T \right) = \frac{\mu_{\alpha} (m, T) \ |W_{\alpha} (m , T )|^2}{\int_{0}^{\infty} dT \ \mu_{\alpha} (m, T) \ |W_{\alpha} (m, T)|^2} \ ,
\end{equation}
which is interpreted as the conditional probability for measuring $T$ at a given mass $m$. The coefficient $\mu$ provides the measure for the identity resolution of the extrinsic boundary states. The necessity for this factor was first pointed out in \cite{Oeckl_2018}. It was explicitly computed in \cite{Fabio_PhD_thesis, christodoulou2023geometry} in the twisted geometry parametrization. For a single link, this reads:
\begin{align}
\label{eq:nu_coeff}
\nu_{\alpha} \left( j \right) & = \frac{j^{- \frac{\alpha}{2}} (1 + 2 j)}{64 \pi^{\frac{7}{2}}} \left( e^{- j^{\alpha + 2}} - e^{- j^{\alpha}(1 + j)^2} \right) \ ,
\end{align}
where $j$ is the spin attached to the link. The coefficient $\mu$ in \eqref{eq:cond_prob_dist} is defined as the product of a factor \eqref{eq:nu_coeff} for each boundary link. In the present context, there are just two types of links. That is four angular links with spin $j_0$ and 12 radial links with spins $j_{\pm}$. 

\medskip

We estimate the crossing time as the expectation value of $T$ over the conditional probability distribution \eqref{eq:cond_prob_dist}:
\begin{equation}
\label{eq:crossing_time_estimate}
\tau_{\alpha} \left( m \right) = \int_{0}^{\infty} dT \  T \ P_{\alpha} \left( m | T \right)  \ .
\end{equation}
The crossing time \eqref{eq:crossing_time_estimate} is well-defined since integrating \eqref{eq:cond_prob_dist} along $T$ at fixed $m$ gives a total (conditional) probability of $1$, regardless of the constant factor multiplying the amplitude \eqref{eq:BW_amplitude}.
\subsection{Crossing time estimate in large spins regime}
\label{subsec:crossing_time_large_j}
As discussed in Section \ref{subsec:balancing_spread}, for large values of spins, we can set the cut-off $K$ equal to zero in the sum over spins in \eqref{eq:extrinsic_coh_state_approx}. From \eqref{eq:nu_coeff} and \eqref{eq:relations_m_T_spins_j0}, it is easy to see that the measure coefficient $\mu$ in the conditional probability distribution \eqref{eq:cond_prob_dist} acts as the Heaviside step function:
\begin{align}
\label{eq:nu_theta}
\mu_{\alpha} \left( m, T \right) \approx \theta \left( T - T_{*} \right) \ ,
\end{align}
where $T_{*}$ is large enough so that $j_0$ in \eqref{eq:relations_m_T_spins_j0} is rounded to the lowest half-integer. Focusing on the first period of the amplitude in $T$, along with condition \eqref{eq:nu_theta}, this results in:
\begin{equation}
\label{eq:P_theta}
P_{\alpha} \left( m | T \right) \approx 
    \begin{cases}
    0 \hspace{23mm} \textrm{for} \hspace{2mm} T \in \left[ 0, T_{*} \right] \\
    \left( \frac{4 \pi m}{ \gamma} - T_{*} \right)^{-1} \hspace{2mm} \textrm{for} \hspace{2mm} T \in \left[ T_{*},  \frac{4 \pi m}{ \gamma} \right] 
    \end{cases} .
\end{equation}
It is immediate to compute the crossing time \eqref{eq:crossing_time_estimate} with \eqref{eq:P_theta}:
\begin{equation}
\label{eq:crossing_time_large_j}
\tau \left( m \right) \approx \frac{m^2 \pi^2}{\gamma^2 \left( \frac{4 \pi m}{ \gamma} - T_{*} \right)} \ .
\end{equation}
We have recovered the linear scaling of the crossing time as a function of the mass $m$. If we completely neglect the $T$ dependence in the boundary data \eqref{eq:relations_m_T_spins_j0}, this is equivalent to consider $T_{*} = 0$, which implies:
\begin{equation}
\label{eq:crossing_time_T*_0}
\tau \left( m \right) \approx \frac{2 \pi m}{\gamma} \hspace{5mm} \textrm{for} \hspace{2mm} T_{*} = 0 \ .
\end{equation}
This is the same estimate originally obtained in \cite{Characteristics_time_scales, christodoulou2023geometry} in the semi-classical regime with $\alpha = - \frac{1}{2}$ using the stationary phase approximation for the amplitude. These two results considered together emphasize that the scaling of the crossing time as a function of $T$ does not depend on the $\alpha$ parameter, which is used to peak to coherent states as discussed in Section \ref{subsec:balancing_spread}. Using the range $\alpha > 0$ makes the calculation remarkably simpler. Since such a regime implies that the extrinsic curvature becomes more and more spread as the spin increases, we infer that the scaling of the crossing time only depends on the intrinsic geometry of the black-to-white hole scenario rather than the extrinsic geometry.
\subsection{The crossing time computation}
\begin{figure}[!hbtp]
\centering
\begin{subfigure}[t]{.49\textwidth}
\includegraphics[width=\linewidth]{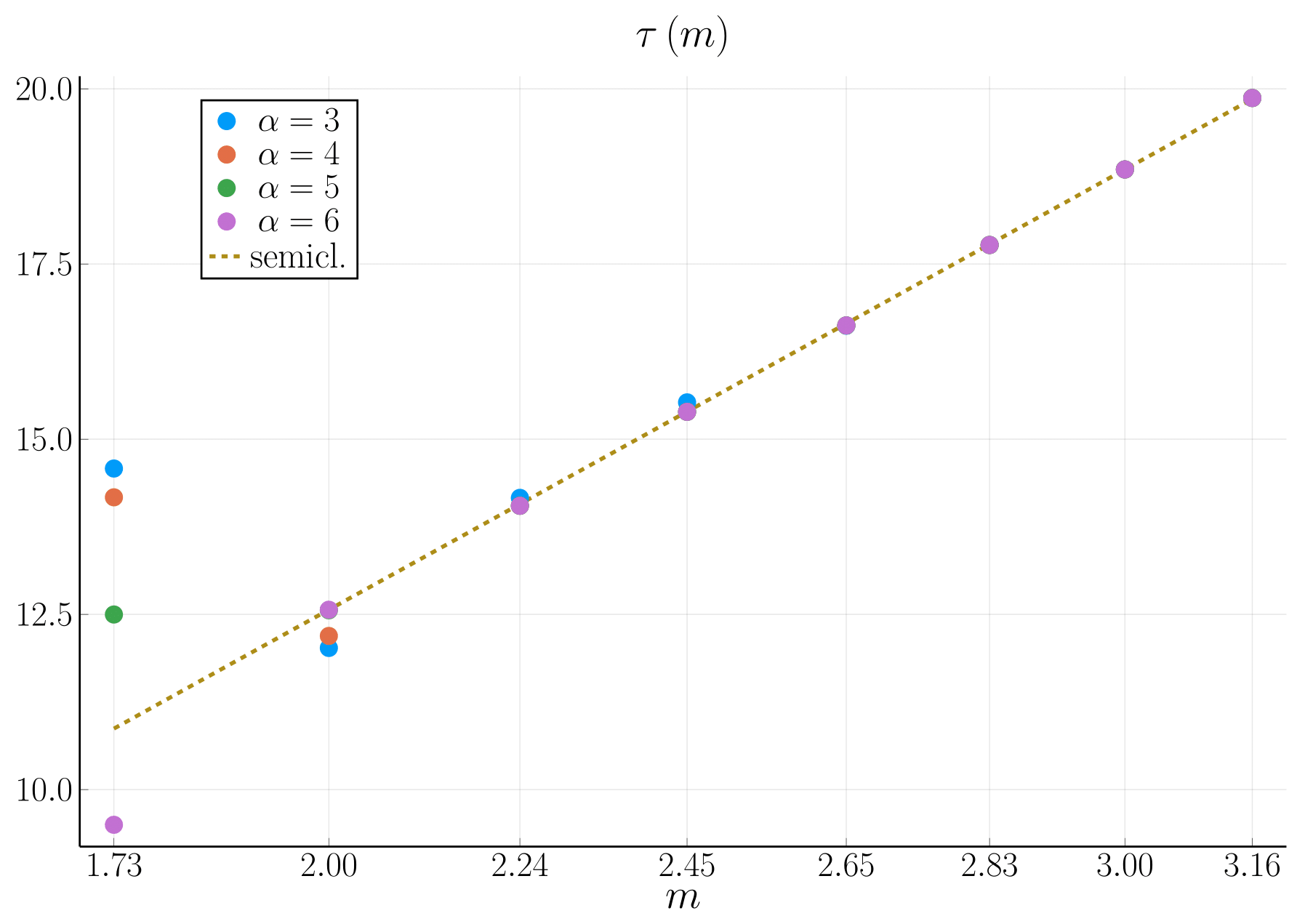}
\end{subfigure} 
\begin{subfigure}[t]{.49\textwidth}
\includegraphics[width=\linewidth]{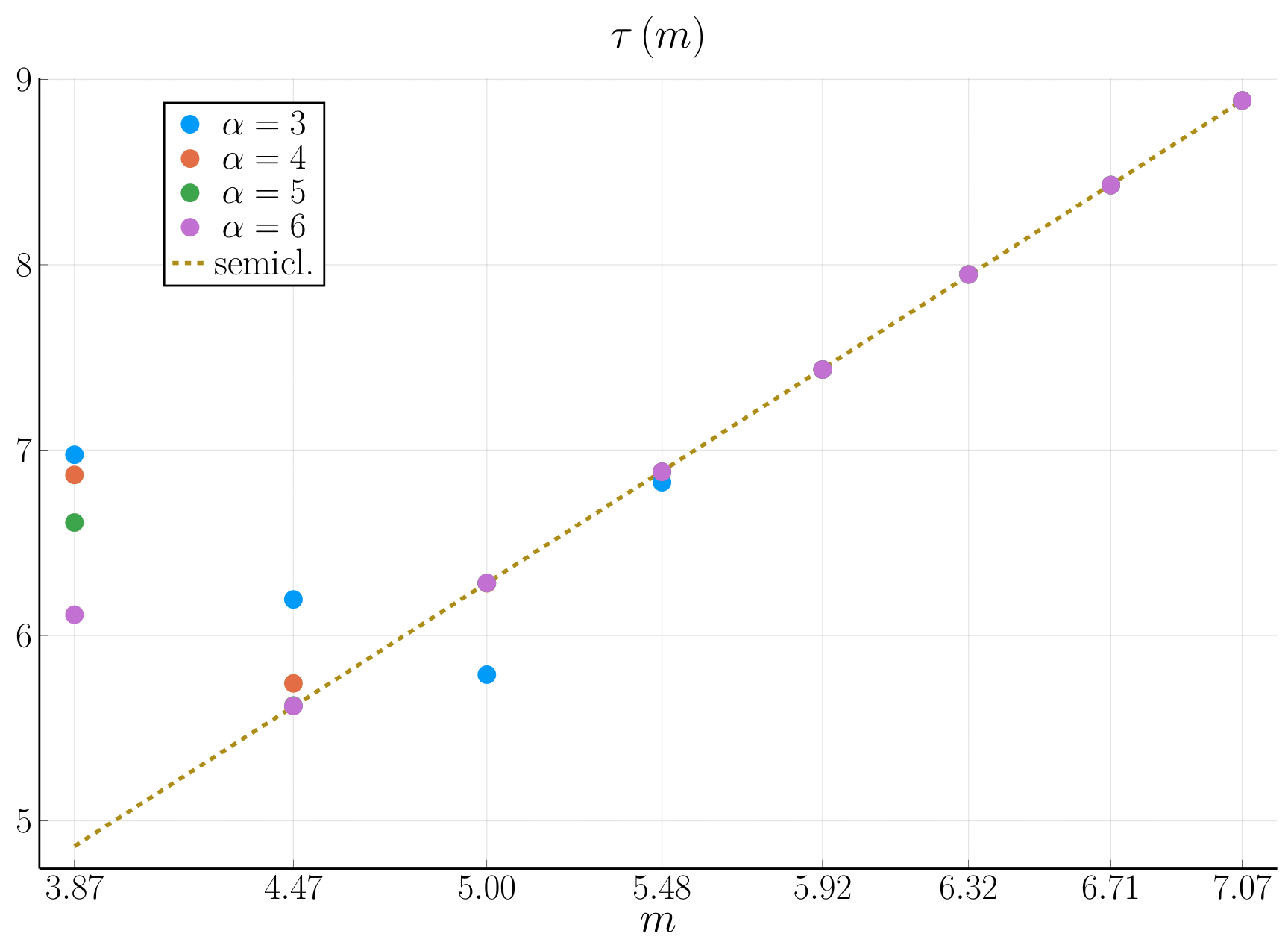}
\end{subfigure} 
\caption{\label{fig:crossing_time}Crossing time \eqref{eq:crossing_time_estimate} evaluated with the amplitudes \eqref{eq:W_partition}, computed with the algorithm discussed in Section \ref{sec:algorithm}. The asymptotic estimate corresponds to \eqref{eq:crossing_time_T*_0}. \textbf{Top:} case $\gamma = 1$. \textbf{Bottom:} case $\gamma = 5$.}
\end{figure}

Finally, we compute the crossing time \eqref{fig:crossing_time} using the numerical approach described in Section \ref{sec:algorithm}. We consider the parameters $\gamma = 1, 5$, $\alpha = 3, \dots 6$,  $K_{0} = 0.5$, $K_{\pm}= 0.5$, $j_{0}^{min}=1.5$, $j_{0}^{max}=5$, $\Delta l^{max}=10$, and $N=100$ for the algorithm described in Section \ref{sec:algorithm}. This choice of parameters is such that approximation \eqref{eq:extrinsic_coh_state_approx} is reliable. After obtaining the amplitudes \eqref{eq:W_partition}, the conditional probability distribution and the crossing time \eqref{eq:crossing_time_estimate} can be evaluated using the trapezoidal rule to compute the integrals over $T$. The result is shown in Figure \ref{fig:crossing_time} for different parameter $\alpha$ values, which balances the quantum spread of the boundary states as discussed in Section \ref{subsec:balancing_spread}. The dashed curve represents the "semiclassical" (in the sense of just the intrinsic geometry) asymptotic estimate \eqref{eq:crossing_time_T*_0}.

\medskip

The crossing time tends to become closer and closer to the asymptotic estimate as a function of $m$ as $\alpha$ increases. The quantum fluctuations emerge for very small values of the black hole mass.  
\section{Conclusions}

In this paper, we presented and described an algorithm to calculate the full black-to-white hole transition amplitude numerically, using the covariant "spinfoam formulation" in the Lorentzian EPRL model and high-performance computing methods. We considered the triangulation originally introduced in \cite{article:Christodoulou_Rovelli_Speziale_Vilensky_2016_planck_star_tunneling_time}, for which a complete numerical evaluation was still missing in the literature. We explicitly applied the algorithm to compute a relevant physical observable corresponding to the crossing time of the transition. We also discussed a straightforward analytical approach to estimate the same quantity alternative to the one currently present in the spinfoam literature, which is based on the stationary phase technique \cite{Fabio_PhD_thesis, christodoulou2023geometry, Characteristics_time_scales}.

\medskip

Compared to the analytical calculation in the literature, we tuned the boundary state in this paper so that the limit for large spins corresponds to an infinite spread of the extrinsic curvature. The estimate of the crossing time (analytical and numerical) performed in this paper shows that the estimate is the same. Physically, this emphasizes that the crossing time of the black-to-white hole transition amplitude does not depend on extrinsic curvature. On the contrary, it appears to be a feature of intrinsic geometry. Therefore, this result adds new information despite being in excellent agreement with the previous estimates of the same physical observable in the literature, including estimates obtained by different communities \cite{Ambrus2005, Barcel__2017}.

\medskip

This work hopes to be the first step in connecting the usage of high-performance computing techniques in loop quantum gravity with the study of the quantum tunneling process between a black hole and a white hole. In recent years, remarkable advances have been made in developing computational methods for spinfoam calculations to study refined triangulation \cite{VertexRenormalizationPaper, Frisoni_2023, han2023complex}. Numerical approaches based on (Markov Chain) Monte Carlo methods combined with the vertex decomposition \eqref{eq:EPRL_vertex_amplitude} recently allowed computing EPRL spinfoam amplitudes and observables for highly non-trivial triangulations, potentially containing infrared bubbles \cite{Radiative_corrections_paper, Self_energy_paper}. Two examples are the vertex renormalization (or "5-1 Pachner move") amplitude \cite{VertexRenormalizationPaper} and the star model \cite{Markov_chain_paper}. The former contains a bubble with $10$ internal faces. We hope that our work will provide a valuable and encouraging ground for progressing in the development of spinfoams more refined than the one considered in this paper.  At the same time, we hope that the considerations about the independence of the crossing time scaling from the extrinsic geometry will encourage the investigations of new probabilistic interpretations of the tunneling process. 
\paragraph*{\bf Acknowledgments}
We thank Pietro Dona for numerous discussions and comments regarding this project, especially for his help in understanding how to associate the normals between tetrahedra of the boundary triangulation. We thank Carlo Rovelli for his continuous support, encouragement, and many enlightening suggestions about this draft. A special thank goes to my Ph.D. advisor Francesca Vidotto. We also thank Fabio D'Ambrosio, Farshid Soltani, Francesco Gozzini, and Muxin Han for their fruitful comments and discussions. We acknowledge the Shared Hierarchical Academic Research Computing Network (SHARCNET) for granting access to their high-performance computing resources. This work was supported by the Natural Science and Engineering Council of Canada (NSERC) through the Discovery Grant "Loop Quantum Gravity: from Computation to Phenomenology." We also acknowledge support from the QISS JFT grant 61466. Western University and Perimeter Institute are located in the traditional lands of Anishinaabek, Haudenosaunee, L\=unaap\`eewak, Attawandaron, and Neutral peoples. %
\centerline{***}

\appendix
\section{Wigner symbols}
\label{app:SU(2)_symbols}
In this paper, we use the definition of the 3j Wigner symbol provided in \cite{book:varshalovic}. It has the following orthogonality property:
\begin{equation}
\sum_{m_1,m_2}\!\! \Wthree{ j_1}{ j_2}{ j_3} {m_1}{m_2}{m_3}\!\! \Wthree{ j_1}{ j_2}{ j_3} {m_1}{m_2}{n_3} = \frac{\delta_{j_{3}l_{3}}\delta_{m_3n_3}}{2 j_3 + 1} \ . 
\end{equation}
The 3j Wigner symbol vanishes if triangular inequalities are not satisfied. We define the 4jm Wigner symbol as the contraction of two 3j symbols over an internal spin $k$:
\begin{align}
\label{eq:4jm_symbol}
& \Wfour{ j_1}{ j_2}{ j_3} {j_4}{m_1}{m_2}{m_3}{m_4}{k} = \\[1em]  \nn
& = \sum_{m_i} (-1)^{k-m_i} \Wthree{ j_1}{ j_2}{ k} {m_1}{m_2}{m_i} \Wthree{ k}{ j_3}{ j_4} {-m_i}{m_3}{m_4} \ .
\end{align}
We also use the synthetic notation:
\begin{equation}
\left(\begin{array}{c} j_f \\ m_f \end{array}\right)^{(k)} \equiv \Wfour{ j_1}{ j_2}{ j_3} {j_4}{m_1}{m_2}{m_3}{m_4}{k} \ . 
\end{equation}
With the definitions \eqref{eq:4jm_symbol} and \eqref{eq:wigner_matrix} we have:
\begin{align}
\label{intD3App}
& \int dn\ D^{j_1}_{m_1, n_1}  \left( n \right) D^{j_2}_{m_2, n_2} \left( n \right) D^{j_3}_{m_3, n_3} \left( n \right) D^{j_4}_{m_4, n_4} \left( n \right) \nn \\[1em] &= 
\sum_{k} d_{k} \Wfour{j_1}{j_2}{j_3}{j_4}{m_1}{m_2}{m_3}{m_4}{k} \Wfour{j_1}{j_2}{j_3}{j_4}{n_1}{n_2}{n_3}{n_4}{k}.
\end{align}
A useful property of Wigner matrices is the following:
\begin{equation}
 D^{j}_{j, m}  \left( n^{-1} \right) = (-1)^{j-m}  D^{j}_{-m, -j}  \left( n \right)  \ .
\end{equation}
We use the irreducible 15j Wigner symbol of the first kind, following the conventions of \cite{GraphMethods}. Its definition in terms of Wigner's 6j symbols turns out to be:
\begin{widetext}
\begin{align}
\label{eq:15jconv}
\{ 15j \} =~  (-1)^{\sum_{i=1}^5 j_i + l_i +k_i} \sum_s d_s \Wsix{j_1}{k_1}{s}{k_2}{j_2}{l_1} \Wsix{j_2}{k_2}{s}{k_3}{j_3}{l_2} 
\Wsix{j_3}{k_3}{s}{k_4}{j_4}{l_3} \Wsix{j_4}{k_4}{s}{k_5}{j_5}{l_4} \Wsix{j_5}{k_5}{s}{j_1}{k_1}{l_5}  \ .
\end{align}
\section{Booster functions} 
\label{app:booster}
The booster functions are the non-compact remnants of the $SL(2, \mathbb{C})$ group \cite{Speziale2016, article:Dona_etal_2019_numerical_study_lorentzian_EPRL_spinfoam_amplitude}. For the physical interpretation of the booster functions and their semiclassical limit, we refer to \cite{dona2020}. We define them as follows:
\begin{align}
\label{eq:boosterdef}
B_4 & \left( j_f, l_f ; i ,  k \right) \equiv \frac{1}{4\pi} \sum_{ p_f } \left(\begin{array}{c} j_f \\ p_f \end{array}\right)^{(i)} \left(\int_0^\infty \mathrm{d} r \sinh^2r \, 
\prod_{f=1}^4 d^{(\gamma j_f,j_f)}_{j_f l_f p_f}(r) \right)
\left(\begin{array}{c} l_f \\ p_f \end{array}\right)^{(k)} \ ,
\end{align}
where $\gamma$ is the Barbero-Immirzi parameter and $d^{(\rho,k)}(r)$ are the matrix elements for $\gamma$-simple irreducible representations of $SL(2,\mathbb{C})$. The explicit form of the boost matrix elements can be found in \cite{Ruhl:1970fk, Speziale2016}. We report below the case of simple irreducible representations:
\begin{align}
\label{eq:dsmall}
d^{(\gamma j,j)}_{jlp}(r) =&  
(-1)^{\frac{j-l}{2}} \frac{\Gamma\left( j + i \gamma j +1\right)}{\left|\Gamma\left(  j + i \gamma j +1\right)\right|} \frac{\Gamma\left( l - i \gamma j +1\right)}{\left|\Gamma\left(  l - i \gamma j +1\right)\right|} \frac{\sqrt{2j+1}\sqrt{2l+1}}{(j+l+1)!}  
\left[(2j)!(l+j)!(l-j)!\frac{(l+p)!(l-p)!}{(j+p)!(j-p)!}\right]^{1/2} \nn \\
&\ \times e^{-(j-i\gamma j +p+1)r}
\sum_{s} \frac{(-1)^{s} \, e^{- 2 s r} }{s!(l-j-s)!} \, {}_2F_1[l+1-i\gamma j,j+p+1+s,j+l+2,1-e^{-2r}] \ .
\end{align}
\end{widetext}
\bibliographystyle{utcaps}

\end{document}